\documentclass[12pt,preprint]{aastex}

\usepackage{natbib}

\shorttitle{Timing NGC 1851 A}
\shortauthors{Freire, Ransom \& Gupta}

\begin{document}

\title{Timing the Eccentric Binary Millisecond Pulsar in NGC 1851}

\author{Paulo C. C. Freire\altaffilmark{1},
  Scott M. Ransom\altaffilmark{2},
  Yashwant Gupta\altaffilmark{3}}
\altaffiltext{1}{National Astronomy and Ionosphere Center, Arecibo
  Observatory, HC3 Box 53995, PR 00612, USA; {\tt pfreire@naic.edu}}
\altaffiltext{2}{National Radio Astronomy Observatory, 520 Edgemont
  Road, Charlottesville, VA 22903, USA; {\tt sransom@nrao.edu}}
\altaffiltext{3}{National Centre for Radio Astrophysics, P.O. Bag 3,
  Ganeshkhind, Pune 411007, India; {\tt ygupta@ncra.tifr.res.in}}

\begin{abstract}
  We have used the Green Bank Telescope to observe the millisecond
  pulsar PSR~J0514$-$4002A on 43 occasions spread over 2 years. This
  5-ms pulsar is located in the globular cluster NGC~1851; it belongs
  to a binary system and has a highly eccentric ($e\,=\,0.888$) orbit.
  We have obtained a phase-coherent timing solution for this object,
  including very precise position, spin and orbital parameters. The
  pulsar is located 4\farcs6 (about 1.3 core radii) from the center of
  the cluster, and is likely to lie on its more distant half. The
  non-detection of eclipses at superior conjunction can be used, given
  the peculiar geometry of this system, to rule out the possibility of
  an extended companion. We have measured the rate of advance of
  periastron for this binary system to be $\dot{\omega} =
  0.01289(4)^\circ \rm yr^{-1}$, which if due completely to general
  relativity, implies a total system mass of $2.453(14)\,\rm
  M_{\odot}$. Given the known mass function, the pulsar mass has to be
  $<$1.5~M$_{\odot}$ and the mass of the companion has to be
  $>$0.96~M$_{\odot}$, implying that it is a heavy white dwarf. The
  350-MHz flux density of this pulsar varies between 0.2 and 1.4 mJy;
  the origin of these variations is not known.
\end{abstract}

\keywords{binaries: general --- pulsars: general --- pulsars:
individual (PSR~J0514-4002A) --- globular clusters : general ---
globular clusters: individual (NGC~1851)}

\section{Introduction}\label{sec:intro}

PSR~J0514$-$4002A was discovered in a 327-MHz search for
steep-spectrum pulsars in globular clusters (GCs) carried out with the
Giant Metrewave Radio Telescope (GMRT) at Khodad, near Pune, India
(Freire et al. 2004, henceforth Paper I)\nocite{fgri04}. The pulsar is
located in the globular cluster NGC~1851 and we will hereafter
designate it to be NGC~1851A. It has a spin frequency of 200 Hz and is
a member of a binary system, with a massive $\sim\,1\,\rm\,M_{\odot}$
companion and a very eccentric orbit ($e = 0.888$).

In the Galaxy, all MSPs are either found isolated or in
low-eccentricity binary systems with low-mass white dwarfs (WDs).
There are pulsars with eccentric orbits and massive companions, but
their spin periods are tens, not hundreds, of Hz. For these systems,
there is not much time for accretion before the massive companion
becomes a neutron star (see Lorimer, 2005 and references
therein)\nocite{lor05}.

NGC~1851A has hybrid characteristics.  Its fast rotation indicates
that the pulsar itself probably formed in the same manner as a normal
MSP, ending up with a light WD companion in a nearly circular orbit.
However, a subsequent stellar interaction could have disrupted the
original binary system, replacing the lowest mass component (the
original WD) with the present companion star in a highly eccentric
orbit. Such exchange interactions are only likely to happen in
environments with very high stellar densities, like the central
regions of globular clusters.  The same process could also form even
more exotic systems, such as MSP/MSP or MSP/Black Hole binaries.

An alternative scenario for the formation of such binaries is the
nearly head-on collision of a neutron star (possibly even an MSP) with
a giant star \cite[e.g.][]{rs91,lpd+06}; this can form an e~=~0.9
binary system consisting of the MSP and the stripped core of the giant
star. We discuss this possibility at length in section \S
\ref{sec:flux}, since it might provide an explanation for the unusual
flux density variations observed for this system.

In Paper I, we had limited information to constrain the nature of the
companion star, no mass estimates for either the pulsar or the
companion, and no precise pulsar timing solution. In order to correct
these deficiencies, we observed the pulsar on 43 occasions over the
last two years with the Green Bank Telescope, and we present the
results below.

\section{Observations and data reduction}
\label{sec:observations}

\subsection{Data taking}

When observing NGC~1851A, we used mostly the prime-focus receiver PF1
with the 350-MHz feed.  On several occasions, however, we used the PF1
receiver with the 425- and 820-MHz feeds, and twice used S-band
receiver.  The back-end instrument was the Pulsar Spigot (Kaplan et
al.~2005)\nocite{kel+05}. For the observations at frequencies below
1\,GHz, where the pulsar is clearly detectable, we observed a total
bandwidth of 50\,MHz, divided in 1024 channels, with a sampling time
of 81.92$\,\mu$s.  The S-band observations had a bandwidth of
600\,MHz, 768 channels, and 81.92$\,\mu$s sampling.  The data were
dedispersed and folded using the {\tt PRESTO} software
package\footnote{\url{http://www.cv.nrao.edu/$\sim$sransom/presto/}},
resulting in 64-bin pulse profiles ($\sim$77\,$\mu$s per bin)
approximately every 500\,s.

With the pulsar's dispersion measure (DM) of 52.14\,pc\,cm$^{-3}$, the
350\,MHz observations had 492\,$\mu$s of dispersive smearing in each
channel, giving an effective system time resolution of 499\,$\mu$s. At
this frequency, dispersive smearing is the dominant contribution to
the width of the Gaussian-like pulse profile, which is
$\sim$590\,$\mu$s at half-power. At 425 MHz the dispersive smearing is
275\,$\mu$s and the total time resolution is 287\,$\mu$s, compared to
the observed pulse width of $\sim$370\,$\mu$s. Therefore dispersive
smearing is still the dominant contribution to the pulse width at
425\,MHz. These measurements suggest that the use of coherent
dedispersion would significantly improve the timing precision of this
pulsar.

\subsection{Timing}

We cross-correlated the average pulse profiles with a Gaussian
template in the Fourier domain \cite{tay92} to obtain topocentric
times of arrival (TOAs). These were then analyzed with {\tt
  TEMPO}\footnote{\url{http://pulsar.princeton.edu/tempo/}}, using the
DE 405 Solar System ephemeris \cite{sta98b} to model the motion of the
Green Bank Telescope relative to the Solar System Barycenter.

After an initial fit, we obtained a $\chi^2$ per degree of freedom of
1.5. Because there are no unmodeled trends in the residuals (their
distribution is Gaussian), we believe that the routine used to derive
the TOAs is slightly under-estimating the TOA uncertainties. After
increasing the TOA uncertainties by 22\% (a value similar to what we
have seen for isolated MSPs timed with the same software) we obtained
a $\chi^2$ per degree of freedom of 1.0. The resulting timing
parameters are presented in Table \ref{tab:parameters}, the
uncertainties are the 1-$\sigma$ {\tt TEMPO} estimates. The
significance of these timing parameters is discussed below.

The orbital parameters were determined using the Damour \& Deruelle
orbital model \cite{dd85,dd86}. We measured
the longitude of periastron ($\omega$) and the time of periastron
passage ($T_0$) at a time (August 2005) where we have a sequence of
daily observations that cover one full orbit.  That $T_0$ was chosen
as the reference epoch. The relativistic parameter $\gamma$, if
included in the model, is very strongly correlated with the projected
semi-major axis $x$ and $\omega$.  When $\gamma$ eventually becomes
needed by the timing model, the measured $x$ and $\omega$ will not
change by more than 0.007\,s and 0.09$^\circ$, respectively, from
their present values.

\section{Position, period and period derivative}
\label{sec:timing}

In Paper I, the position of the source identified as the pulsar in an
interferometric map of the cluster is given as $\alpha = 05^{\rm
  h}14^{\rm m}06.74\pm0.06^{\rm s}$, $\delta = -40^\circ
02'50.0\pm1.3''$ (J2000). The $\alpha$ and $\delta$ given in Table
\ref{tab:parameters} are 1-$\sigma$ consistent with these values, but
much more precise.  The pulsar is about 4$\farcs$6 (about 1.3 core
radii) East of the center of the cluster (see Table \ref{tab:GC}).
Such close proximity to the center is typical for GC
pulsars\footnote{See, e.g.,
  \url{http://www2.naic.edu/$\sim$pfreire/GCpsr.html}} due to the
effects of mass segregation. The uncertainty in the derived magnitude
of the projected offset from the center of the cluster is dominated by
uncertainties in the position of the cluster center and the cluster's
core radius.

The observed period derivative, $\dot{P}_{\rm obs}$, is the sum of
several different terms \cite{phi92b}:

\begin{equation}
\label{eq:accelerations}
\left( \frac{\dot{P}}{P} \right)_{\rm obs} =
\left( \frac{\dot{P}}{P} \right)_{\rm int} + \frac{1}{c} \left( a_G +
\mu^2 D + a_c \right),
\end{equation}

\noindent where $\dot{P}_{\rm int}$ is the intrinsic period derivative
of the pulsar, $a_G$ is the difference in Galactic accelerations
between the Solar System Barycenter and NGC~1851 projected along the
line-of-sight, $\mu$ and $D$ are the proper motion and distance to the
pulsar (assumed to be the same as the cluster values in Table
\ref{tab:GC}) and $a_c$ is the acceleration of the binary caused by
the cluster, also projected along the line-of-sight. For $a_G$, we
obtain $6.1\,\times\,10^{-12}\rm m\,s^{-2}$ \cite{kg89}. For $\mu^2
D$, we obtain $6.5\,\times\,10^{-11}\rm m\,s^{-2}$, using the value
for $\mu$ obtained by Dinescu et al.~(1999)\nocite{dga99}. With these
values and $\dot{P}_{\rm obs}$ we obtain:

\begin{equation}
\label{eq:acc2}
\dot{P}_c \equiv \dot{P}_{\rm int} + P \frac{a_c}{c}
= (+0.0\,\pm\,1.4)\,\times\,10^{-22}. 
\end{equation}

%
Pulsars with negative $\dot{P}_c$ can put lower
limits on the absolute value of $a_c$, and for that reason have been
used to study the gravitational potential of their parent GCs (e.g.,
Possenti et al.  2002, Freire et al. 2003)\nocite{dpf+02,fck+03}. In
this case, that can't be done, but we can say that it is very likely
that the pulsar is located on the more distant half of the cluster.
If it were on the near side, then $a_c$ in eqn. \ref{eq:acc2} would be
positive, and $\dot{P}_{\rm int} \,< 2.8\,\times\,10^{-22}$
(2$-\sigma$ upper limit). This would imply an anomalous characteristic
age of more than 280 Gyr. In the likely event that the pulsar has a
more normal characteristic age, then $a_c$ has to be negative, placing
the pulsar on the far side of the cluster.

In Table \ref{tab:GC}, $v_z(0)$ is the spread of stellar line-of-sight
velocities at the core. Using this, the cluster core radius
and an analytical gravitational model for the central regions of 
GCs \cite{fhn+05}, we can calculate the maximum negative acceleration
that the cluster can induce on the pulsar at its sky position. From
this we can derive an upper limit on the intrinsic period derivative
and surface magnetic flux density, and a lower limit on the
characteristic age (see Table \ref{tab:parameters}), as assumed 
in the formation scenarios in Paper I.

\section{On the nature of the companion}
\label{sec:companion}

\subsection{Is the companion extended?}

In Paper I, we reached the conclusion that, in the event of an
extended companion, the circularization timescale for this system is
larger than the age of the cluster. An extended companion could not
therefore be disproved because of the system's eccentricity.

In Table \ref{tab:geometry}, we present results of calculations of the
companion masses, absolute component separation at superior
conjunction ($a_{\rm sup}$) and its projection in the plane of the sky
($S = a_{\rm sup} \cos i$) for a set of orbital inclinations. The
companion masses are calculated for a pulsar mass of 1.2~M$_{\odot}$;
the mass of the lightest known NS, the companion of PSR~J1756$-$2251
(Faulkner et al.~2005, Ferdman et al.~2006)\nocite{fkl+05,fsb+06}.
For these companion masses, we list the radii of zero-age
main-sequence stars, calculated using $\log_{10} (R / R_{\odot}) =
0.917 \times \log_{10} (M / M_{\odot})-0.020$, which apply for stars
of mass similar to that of the Sun \cite{ac00}. The use of the
smallest possible pulsar mass and corresponding zero age main sequence
radii leads to the lowest possible companion masses and (if extended)
sizes, the most conservative assumption in the discussion that follows.

For $i < 60^\circ$ (limits are approximate), the companion masses are above
1\,M$_{\odot}$. In a star cluster as old as NGC~1851 (about 9\,Gyr,
see Tab. \ref{tab:GC}), such stars have either finished their life
cycles and left compact remnants behind, or are now in their giant
phases.  Giant stars have radii that are tens to hundreds of times that
of similarly massive stars on the main sequence, which are of the
order of one solar radius. Such giant stars would not even fit within
the distance between the components of NGC~1851A at superior
conjunction or periastron. In the first scan taken in 2006 July 7, we
can detect the pulsar clearly through periastron, implying that the
possibility of such a giant companion is excluded.

For $i > 80^\circ$, the radius of a main-sequence companion star is
larger than $S$, implying that an eclipse should occur. The longitude
of periastron for this system is 82.27$^\circ$, and therefore superior
conjunction occurs at a true anomaly of 7.73$^\circ$, 15\,min and
54.1\,s after periastron. For the second scan taken on 2006 July 7,
the two TOAs occur 11.3 and 22.5 min of barycentric time after the
periastron\footnote{Because of the advance of periastron between the
epoch of the parameters in Table \ref{tab:parameters} and this
observation, superior conjunction occurred 15\,min and 52.8\,s after
periastron, a difference of 1.3\,s relative to the reference
epoch.}. The pulsar is detectable throughout the whole scan. This
excludes the possibility of the companion being a main sequence star
with a mass smaller than 0.87~M$_{\odot}$.

For $60^\circ < i < 80^\circ$, we could in principle avoid eclipses
and still postulate a main-sequence companion. For any higher pulsar
mass, or extended stars larger than zero age main-sequence, this inclination
interval becomes smaller. However, even for the lowest inclination
listed in Table \ref{tab:geometry}, the pulsar signal would be passing
only 1.2~R$_{\odot}$ above the surface of the companion, right through
its corona. We also note that the pulsar wind should lead to a denser
and more extended corona in this star, so we should be able to detect some
increase in the electron
column density at this orbital phase. Dividing the second observation
of 2006 July 7 in 4 sub-bands and calculating TOAs for the whole
observation, we obtain a DM of 52.148(5)\,pc\,cm$^{-3}$, not
measurably different than the DM at other orbital phases and smaller
than the increases in the plasma column density seen at superior
conjunction for pulsars with extended companions (see Freire,
2005\nocite{fre05} and references therein).

Furthermore, if the companion was extended, there would be tidal
effects on the orbit of the pulsar (e.g., PSR~J0045$-$7319, Lai 1996 and
PSR~B1259$-$63, Wang et al. 2004)\nocite{lai96,wjm04}. These include
changes in the inclination (and therefore, changes in the projected
size $\dot{x} \equiv a_{p}/c\, d(\sin i)/dt$) of the orbit, large
variations in the orbital period ($\dot{P_B} \neq 0$), and unmodeled
and systematic trends in the residuals. As described above, the values
for $\dot{x}$ and $\dot{P_B}$ are not significant (see Table
\ref{tab:parameters}), and the object times as well as an isolated
MSP, with no unmodeled trends in the residuals despite the fact
that our timing precision is much higher than for either PSR~J0045$-$7319
or PSR~B1259$-$63. In order to avoid eclipses and any measurable tidal
effects, the companion has to be a compact object. The implications of
this are discussed below.

\subsection{What is the mass of the companion?}

For a system consisting of two compact objects, the observed
$\dot{\omega}$ is due solely to the effects of general relativity.
This allows an estimate of the total mass of the system:
$2.453(14)\,\rm M_{\odot}$. Given the mass function, the pulsar mass
cannot be larger than 1.50\,M$_{\odot}$ and the companion mass must be
larger than 0.96\,M$_{\odot}$. For a median inclination of 60$^\circ$,
the mass of the pulsar is 1.350\,M$_{\odot}$, a value that is fairly
typical of the neutron stars with well-determined masses. In this
case, the companion mass would be about 1.105\,M$_{\odot}$.

\begin{figure}
\plotone{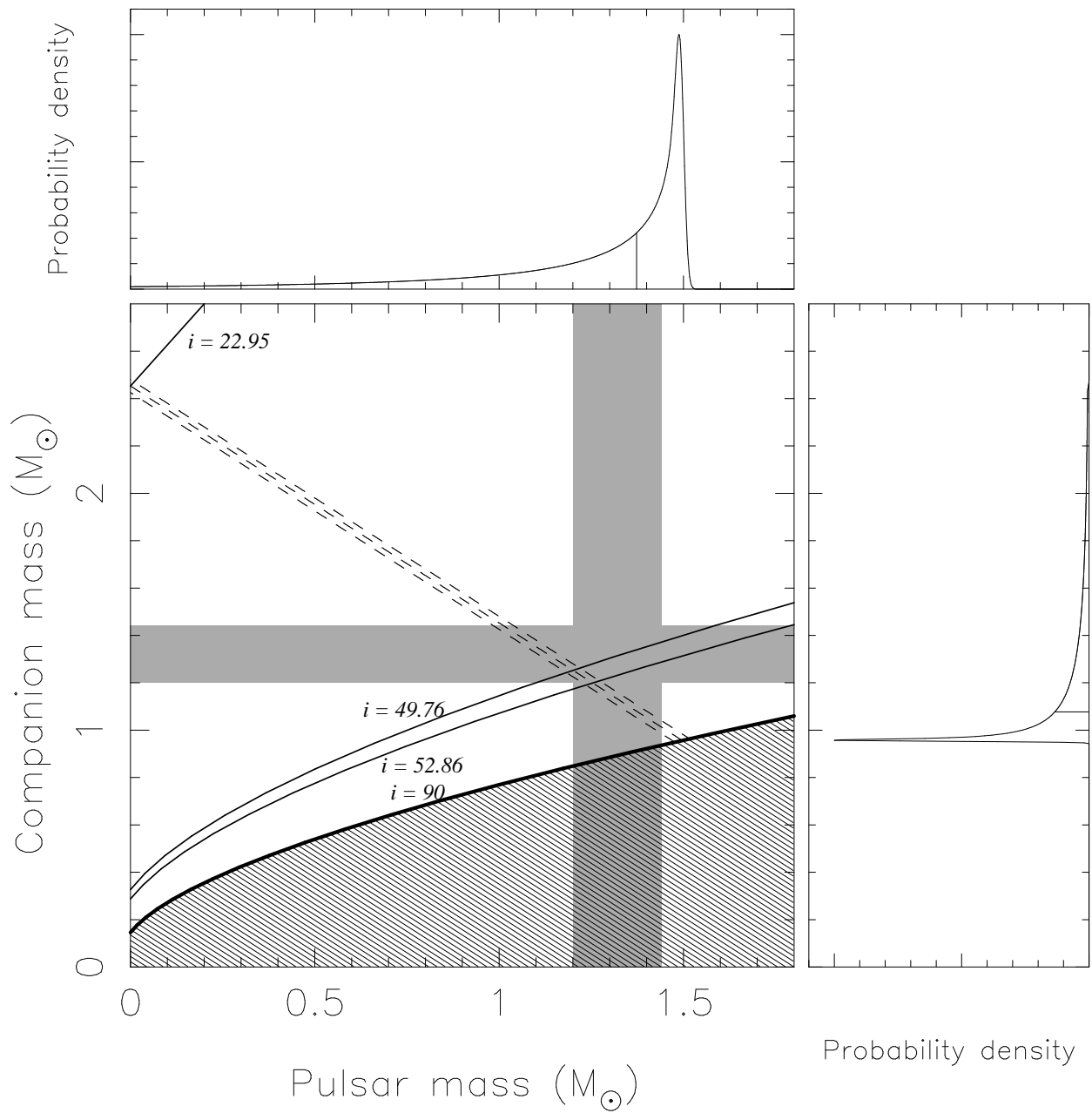}
\caption{\label{fig:mass_mass} Constraints on the masses of NGC 1851A
  and its companion. The hatched region is excluded by knowledge of
  the mass function and by $\sin i \leq 1$.  The diagonal dashed lines
  correspond to a total system mass that causes a general-relativistic
  $\dot{\omega}$ equal or within 2-$\sigma$ of the measured value.
  The four solid curves indicate constant inclinations. We also
  display the probability density function for
  the mass of the pulsar ({\em top}) and the mass of the companion
  ({\em right}), and mark the respective medians with vertical
  (horizontal) lines.
}
\end{figure}

For $49.76^\circ < i < 52.86^\circ$, both components would have masses
within the present range of well-measured neutron stars (see Fig.
\ref{fig:mass_mass}); however, given a flat probability distribution
in $\cos i$, it is about 15 times more likely that $52.86^\circ < i <
90^\circ$, where $m_c < 1.2\,\rm M_{\odot}$. In the higher inclination
range, there is the possibility that the companion is the lightest NS
ever discovered.  However, this is probably more than compensated for
in the lower inclination range by the possibility that the companion
is a WD or a stripped core of a giant star with $m_c > 1.2\,\rm
M_{\odot}$. We therefore believe that the probabilities are not far
from 15 to 1 against the companion being a neutron star.  Given the
slight possibility, though, that the companion star could be a MSP, we
searched several of the 350, 820, and 1950\,MHz observations for
additional pulsations, but found none.

\section{Flux Density Variations}
\label{sec:flux}

In Table \ref{tab:parameters} we list the average flux densities at
350, 425, 820 and 1950\,MHz, together with the number of observations
averaged, we derive a spectral index of $-$2.65. At 350\,MHz, the flux
density changes very significantly, from about 0.2 to 1.4\,mJy. Even
our brightest detection is significantly fainter than the average flux
density reported in Paper I (3.4$\,\pm\,$0.4\,mJy).  This could
explain the discrepancy in reported spectral indices; in Paper I we
use the non-detection at 610\,MHz to derive $\alpha_S\,<\,-3.4$.
However, the GMRT flux density at 325\,MHz comes from imaging, so it
is possible that there is extra unpulsed radio emission in the
vicinity of the pulsar. However, we can not exclude the possibility of
some lapse in our understanding of the sensitivity of these systems.

The timescale for the variations near periastron ($\tau_p$) is of the
order of 1000\,s; near apastron the flux density is seen to change
very slowly and monotonically over several hours, only becoming
significant from one day to the next (i.e.~$\tau_a/\tau_p > 10$).

The flux density variations are almost certainly not caused by
diffractive scintillation.  The scintillation bandwidth at 350\,MHz is
(very roughly) 10$-$20\,kHz (Cordes and Lazio 2002), so our
observations average over many hundreds of scintles in the 50\,MHz
bandwidth.  Furthermore, refractive scintillation will change the flux by
no more than $\sim$50\%, on timescales of $\sim$10$^7$ seconds (Cordes
and Lazio, 2002). Therefore, if the ISM between us and the cluster
has everywhere a Kolmogorov spatial spectrum and is described, to
within a few orders of magnitude, by the Cordes and Lazio model, then
no large-amplitude variations in the flux density should be
observed. Scintillation by a region or regions of particularly high
ionized gas density could provide an explanation, we could
alternatively be dealing with small eclipses caused by gas in the
vicinity of the binary.

\subsection{Gas from the Companion Star}

Flux density variations of similar magnitude and duration have been
observed for several eclipsing binary pulsars away from superior
conjunction.  The best documented examples are PSR~J1740$-$5340, a
MSP-main sequence binary located in the GC NGC~6397 (Ferraro et al.
2001)\nocite{fpds01}, and PSR~J1748$-$2446A, an eclipsing binary MSP
located in the GC Terzan~5 \cite{ljm+90,lmbm00} also known as
Terzan~5~A. As in virtually all eclipsing binaries, the companions of
these two pulsars are losing gas as a result of their interaction with
the pulsar wind. Part of the observed flux density variations away
from superior conjunction are due to secondary eclipses thought to be
caused by gas clumps that remain in the vicinity of these binary
systems; even though the precise mechanism of the modulation is not
known. Many, but not all, of these secondary eclipses are ``total''
(i.e.~the pulsars become completely undetectable).

The companion of NGC~1851A is much more massive than the companions of
either PSR~J1740$-$5340 or Terzan~5~A. No outgassing has ever been
clearly detected by pulsar timing in systems where the companion is a
massive white dwarf, and it has not been detected in NGC~1851A either
(\S \ref{sec:companion}). However, there could still be gas clumps in
the vicinity of the system if the companion of NGC~1851A is outgassing
at a lower level than what we can detect. This is a possibility if it
is a stripped core of a giant star.

Unlike PSR~J1740$-$5340 or Terzan~5~A, in NGC~1851A we detect no
``total'' eclipses caused by gas clumps at superior conjunction or
away from it.  Furthermore, a common feature of such eclipses is that
they invariably affect the lower frequencies more strongly
(e.g.~longer duration eclipses), although in varying degrees for
different binary systems. That is not observed in NGC~1851A, where the
attenuation may be frequency dependent, although in a variable and
inconsistent manner.  However, it is important to note that the
behavior of the other pulsars mentioned above varies strongly from
orbit to orbit. In our case, we only have good observations through a
single periastron and superior conjunction.  With the current data we
cannot conclusively demonstrate that the variations are caused by
orbiting gas.  However, we do consider it a possible explanation for
the observed flux density variations.

\subsection{Refractive scintillation}

It is possible that, despite the predictions of Cordes and Lazio, the
changes in flux density are indeed due to scintillation. This is more
likely if, somewhere between us and the pulsar, there are unusual
regions of the ISM acting as strong lenses. We can use
the changes in $\tau$ to constrain their location.
Assuming a maximum inclination of 90$^\circ$, the pulsar is
traveling at $v_p\sim$l73\,km\,s$^{-1}$ relative to the center of mass
at periastron, and at $v_a\sim$10\,km\,s$^{-1}$ relative to the center
of mass at apastron. Furthermore, the distance and proper motion
listed for this cluster in Table \ref{tab:GC} imply a perpendicular
cluster velocity of $\sim$160~km~s$^{-1}$.

If the lenses are near us, at apastron the velocity of the pulsar
$v_a'$ relative to those lenses can not be much different from the
perpendicular velocity of the cluster times $f$ (the distance from the
Solar System to the lens divided by the distance from the Solar System
to NGC~1851). At periastron, we obtain $(20 \times f < v_p' < 320
\times f) $~km~s$^{-1}$, depending on how the pulsar's orbital
velocity at periastron adds to its proper motion. This would imply
$0.1 < v_p' / v_a' = \tau_a/\tau_p < 2$, which is inconsistent with
observation. If $f$ is very small, then the Earth's orbital velocity
becomes dominant in determining $\tau$ and we can then produce a
larger range of $\tau_a/\tau_p$, however, no seasonal variations in
$\tau_a$ are observable. If, on the other hand, the medium responsible
for the variations has a small velocity relative to the center of mass
of the binary, we obtain $\tau_a/\tau_p \sim f v_p / f v_a =
(1+e)/(1-e) = 16.85$, which is consistent with observation.

This suggests that a hypothetical lensing medium would probably be
located in the cluster, which is also the case if we were instead
observing gas clumps orbiting the binary itself. Given the proximity
to the pulsar, we can then readily estimate the associated spatial
scale of the medium, independently of its nature, from $v_p$ and
$\tau_p$: about 10$^5$ km.

\section{Conclusion and prospects}
\label{sec:discussion}

In this work we have determined the phase-coherent timing solution of
NGC~1851A. This allowed us to locate the pulsar in the globular
cluster and determine precise rotational and orbital parameters. The
geometry of the system, the lack of any observable tidal effects in
the timing and the lack of eclipses or a measurable increase of the DM
at superior conjunction allow us to exclude the possibility of an
extended companion. This in turn allows us to calculate the total mass
of the system from the observed rate of the advance of periastron:
2.453(14)\,M$_{\odot}$. The pulsar mass has to be smaller than
1.5~M$_{\odot}$ and the companion mass has to be larger than
0.96~M$_{\odot}$. The companion is either a massive WD or the stripped
core of a giant star, and less likely a neutron star.

The flux density variations remain puzzling. They could be due to gas
clumps in the vicinity of the binary created by outgassing from the
companion, a possibility considering that it could be the stripped
core of a giant star. Alternatively, they could be due to refractive
scintillation, in which case the lensing structures are also likely to
be located in NGC~1851. This would imply the presence of intra-cluster
gas, the second instance after 47~Tucanae \cite{fkl+01} where such gas
was confirmed in a globular cluster.

If we observe the pulsar often close to periastron, it might be
possible to measure the velocity of the binary relative to the medium
that causes the flux density variations and decide which of the
scenarios discussed is the correct one. Furthermore, a detailed study
of the orbital variability of
$\tau$ might allow a determination of the orbital inclination of the
system, and lead to a determination of the individual masses. So far,
only one MSP, PSR~J1909$-$3744 has a well-determined mass,
1.44\,M$_{\odot}$ \cite{jhb+05}.  Measuring more MSP masses precisely
is important in order to estimate how much mass a neutron star needs
to accrete in order to become a MSP. 

Such a detailed mass measurement will be confirmed (or refuted) by the
measurement of another relativistic timing effect, the Einstein delay
($\gamma$). Our simulations suggest that, with the present observing
strategy and timing precision, we will be able to measure $\gamma$ to
better than 1 ms within 6 to 8 years. For an edge-on orbit, $\gamma$
is predicted to be 16.13\,ms, and for $i\,=\,60^\circ$, $\gamma \,= \,
19.44\,$ms. The Shapiro delay can not be measured with the present
timing precision, no tests of general relativity are possible with
the present telescope sensitivities.

\acknowledgements

The National Radio Astronomy Observatory is a facility of the National
Science Foundation operated under cooperative agreement by Associated
Universities, Incorporated. We thank Jason Hessels and Fernando
Camilo for help with observations, David Nice for assistance with TOA
simulation software and Christopher Salter for a first skeptic review
of the paper.

\clearpage

\begin{deluxetable}{ l c l }
\tablecolumns{3}
\tablewidth{0pc}
\tablecaption{Parameters for the globular cluster NGC 1851}
\startdata
\hline
\hline
Right Ascension of center, $\alpha$ & $05^{\rm h}14^{\rm m}06\fs 3$ & \cite{har96}\\
Declination of center, $\delta$ & $-40^\circ 02\arcmin 50$ &
\cite{har96} \\
Galactic longitude, $l$ ($^\circ$) & 244.51 \\
Galactic latitude, $b$ ($^\circ$) & $-$35.04 \\
Cluster distance, $D$ (kpc) & 12.1 & \cite{har96} \\
Core radius, $\theta_c$ (arcmin) & 0.06 & \cite{har96} \\
$v_z(0)$ (km s$^{-1}$)& 11.3 & (Dubath, Meyland \& Mayor
1997)\nocite{dmm97}\\
Proper Motion in $\alpha$, $\mu_{\alpha}$ (mas yr$^{-1}$) & +1.28$\,\pm\,$0.68 &
(Dinescu et al. 1999) \nocite{dga99} \\
Proper Motion in $\delta$, $\mu_{\delta}$ (mas yr$^{-1}$) & +2.39$\,\pm\,$0.65 &
(Dinescu et al. 1999) \nocite{dga99} \\
Age (Gyr) & 8.9$\,\pm\,$0.9 & \cite{sw97}
\enddata
\label{tab:GC}
\end{deluxetable}

\clearpage

\begin{deluxetable}{ l c }
\tabletypesize{\footnotesize}
\tablecolumns{2}
\tablewidth{0pc}
\tablecaption{Parameters for the PSR J0514$-$4002A binary system}
\startdata
\hline
\multicolumn{2}{c}{Observation parameters and flux density} \\
\hline
First observation (MJD) & 53258 \\
Last observation (MJD) & 54066 \\
Number of TOAs & 316 \\
Residual rms ($\mu$s) & 30 \\
Epoch (MJD) & 53623.155088 \\
Flux density, & \\
... at 350 MHz, $S_{350}$ (mJy) & 0.62(2) (93)\tablenotemark{a}\\
... at 425 MHz, $S_{425}$ (mJy) & 0.28(6) (2)\\
... at 820 MHz, $S_{820}$ (mJy) & 0.065(6) (12) \\
... at 1950 MHz, $S_{1950}$ (mJy) & 0.0056(10) (3) \\
Spectral index, $\alpha_S$ & $-$2.71(9) \\
\hline
\multicolumn{2}{c}{Timing parameters} \\
\hline
Right Ascension, $\alpha$ (J2000) & $05^{\rm h}14^{\rm m}06\fs 6927(2)$ \\
Declination, $\delta$ (J2000) & $-40^\circ 02\arcmin 48\farcs 897(2)$
\\
%
%
%
Spin frequency, $\nu$ (Hz)  & 200.37770740529(11) \\
Time derivative of $\nu$, $\dot{\nu}$ (10$^{-17}$ Hz$^2$) & $-$4.7(5) \\
Dispersion Measure, DM (pc\,cm$^{-3}$) & 52.1489(6) \\
Orbital period, $P_b$ (days) & 18.78517915(4) \\
Projected size or orbit, $x$ (l-s) & 36.296588(9) \\
Time of passage through periastron, $T_0$ (MJD) & 53623.1550879(4) \\
Eccentricity, $e$                 & 0.8879773(3) \\
Longitude of periastron, $\omega$ ($^\circ$)  & 82.266550(18) \\
Rate of advance of periastron, $\dot{\omega}$ ($^\circ$ yr$^{-1}$) &
0.01289(4)\\
Einstein delay, $\gamma$ (ms) & [8$\pm$26]\tablenotemark{b}\\
Time derivative of $x$, $\dot{x}$ ($10^{-12}$ l-s/s) &
[$-$0.1$\pm$0.4] \\
Time derivative of $P_B$, $\dot{P_B}$ ($10^{-9}$) & [1.5$\pm$0.7] \\
\hline
\multicolumn{2}{c}{Derived parameters} \\
\hline
Angular distance from center of cluster, $\theta_{\perp}$ & \\
... (arcseconds) & 4.6 \\
... (core radii) & 1.3 \\
... (pc) & 0.27 \\
Spin period, $P$ (ms) & 4.990575114114(3) \\
Time derivative of $P$, $\dot{P}$ (10$^{-21}$) & 1.17(14) \\
Maximum cluster acceleration, $a_{c, \rm max}$ (m.s$^{-2}$) &
18.4$\,\times\,10^{-9}$\\
Maximum intrinsic $\dot{P}$, $\dot{P}_{\rm int, max}$
(10$^{-19}$) & 3.0 \\
Maximum magnetic flux density, $B_0$ ($10^{9}$ Gauss) & 1.2 \\
Minimum characteristic age, $\tau_{c, \rm min}$ (Gyr) & 0.26 \\
Mass function, $f\,(\rm M_{\odot})$ & 0.14549547(11) \\
Total system mass, M $(\rm M_{\odot})$ & 2.453(14) \\
Maximum pulsar mass, M$_{p, \rm max}\,(\rm M_{\odot})$ & 1.50 \\
Minimum companion mass, M$_{c, \rm min}\,(\rm M_{\odot})$ & 0.96 \\
\enddata
\tablenotetext{a}{Number of measurements averaged in parentheses. Flux
uncertainty is 30\% / $\sqrt{N}$}
\tablenotetext{b}{Values in square brackets are not 3-$\sigma$
  significant, nor are they expected to be so. They were not fit when
  determining the remaining timing parameters.
}
\label{tab:parameters}
\end{deluxetable}

\clearpage

\begin{deluxetable}{ l l l l l}
\tablecolumns{5}
\tablewidth{0pc}
\tablecaption{Geometric parameters for the companion of NGC~1851A}
\startdata
\hline
$i$ &  $m_c$ & $a_{sup}$ & $S = a_{\rm sup} \cos i$ & $R_{c}$ \\
($^\circ$) & (M$_{\odot}$) & (R$_{\odot}$) & (R$_{\odot}$) & (R$_{\odot}$)\\
\hline
90 & 0.8483 &   4.2471 & 0       & 0.82 \\

85 & 0.8528 &   4.2502 & 0.3704  & 0.83 \\

80 & 0.8665 &   4.2596 & 0.7397  & 0.84 \\

75 & 0.8901 &   4.2758 & 1.1067  & 0.86 \\

70 & 0.9252 &   4.2996 & 1.4705  & 0.89 \\

65 & 0.9739 &   4.3322 & 1.8309  & 0.93 \\

60 & 1.0396 &   4.3754 & 2.1877  & 0.99 \\

\enddata
\label{tab:geometry}
\end{deluxetable}

\clearpage

\end{document}